\documentclass[a4paper,11pt]{article}
\usepackage{jcappub} % for details on the use of the package, please see the JINST-author-manual
\usepackage{multirow} 

\arxivnumber{2507.09981} % Only if you have one
\title{\boldmath New Insights into Dark Energy from DESI DR2 with CMB and SNIa}
\author[\,a]{Da-Chun Qiang,}
\affiliation[a]{\,Institute for Gravitational Wave Astronomy, Henan Academy of Sciences,\\
Zhengzhou 450046, Henan, China}
\author[\hspace{1.5mm}b]{Jing-Yi Jia}
\author[]{and Hao Wei$^{\hspace{1.5mm}b,*}$}
\affiliation[b]{\,School of Physics, Beijing Institute of Technology, \\
Beijing 100081, China}
\emailAdd{dcqiang@hnas.ac.cn, jjy@bit.edu.cn, haowei@bit.edu.cn}

\abstract{Analyses by the Dark Energy Spectroscopic Instrument (DESI) collaboration suggest a significant deviation from the $\Lambda$CDM model when their baryon acoustic oscillation (BAO) measurements are combined with Planck cosmic microwave background (CMB) data and various Type Ia supernova (SNIa) samples. In this work, we systematically investigate the origin of the deviations from the $\Lambda$CDM reported in recent cosmological analyses by combining different CMB datasets, BAO measurements, and DESY5 SNIa samples within the $w_0w_a$CDM framework. We find that the DESY5 SNIa sample, particularly its low-redshift component (DES-lowz), the Planck CMB data, the lensing measurements of Planck and ACT-DR6, and the DESI-DR2 BAO measurements contribute most significantly to the observed tensions. In contrast, combinations involving DES-SN, WMAP, SPT, and ACT-DR6 remain consistent with $\Lambda$CDM within $\sim1\sigma$. Our results highlight the critical impact of SNIa systematics, CMB data, and the choice of BAO dataset on constraints of dynamical dark energy models. These findings underscore the importance of improved calibration, homogeneity, and cross-validation of observational datasets to robustly assess potential deviations from the standard cosmological model.}

{
\footnotetext[1]{Corresponding Author}
}

\begin{document}
\maketitle
\flushbottom
\section{Introduction}
\label{sec:intro}

The release of Baryon Acoustic Oscillation (BAO) measurements by the Dark Energy Spectroscopic Instrument (DESI) has reignited interest in understanding the nature of dark energy~\cite{desidr1,desidr2}. Combining cosmic microwave background (CMB) data with DESI DR1 BAO measurements and three different Type Ia supernova (SNIa) datasets—Pantheon+~\cite{pantheon+_1,pantheon+_2}, Union3~\cite{union3}, and Dark Energy Survey Year-Five Data Release (DESY5)~\cite{DESY5-1,DESY5-2,DESY5-3}—reveals deviations from the $\Lambda$CDM cosmology within the $w_0w_a$CDM~\cite{cpl-1,cpl-2} framework at significance levels of $2.5\sigma$, $3.5\sigma$, and $3.9\sigma$, respectively~\cite{desidr1}. These tensions increase to $2.8\sigma$, $3.8\sigma$, and $4.2\sigma$ when the DESI BAO data are updated from Data Release 1 (DR1) to Data Release 2 (DR2)~\cite{desidr2}. Even without any SNIa datasets, the combined DESI DR2 BAO and CMB data exhibit a $3.1\sigma$ tension with $\Lambda$CDM cosmology~\cite{desidr2}. In response to these tensions, significant efforts have been devoted to exploring new physics scenarios (theoretical models), such as non-standard dark matter evolution~\cite[e.g.,][]{dm1,dm2,dm3,dm4,dm5,dm6}, ultra-light axion-like fields~\cite{2020PhRvL.125v1301M,2022NatRP...4..452K} as dynamical dark energy candidates~\cite[e.g.,][]{2025arXiv250318120N,2025arXiv250318417L,2025arXiv250318924N,2025arXiv250320178U,2025arXiv250417638L}, and potential interactions between dark matter and dark energy~\cite[e.g.,][]{2024PhRvL.133y1003G,2024arXiv241209064L,2025arXiv250310806C,2025PDU....4801935F,2025arXiv250316415K,2025arXiv250618477L,2025arXiv250321652S,2025PhRvD.111l3511S,2025arXiv250400985Y,2025arXiv250400994P}.

Alongside the development of theoretical models, considerable effort has also been devoted to the detailed analysis of observational data (CMB, BAO, and SNIa)~\cite[e.g.][]{William-CMB_cpl,huang_H0_CMB,2025SCPMA..6800413H,2025MNRAS.538..875E,2024JCAP...12..007C,2025arXiv250103480P,2025MNRAS.540.1626D,2024ApJ...976L..11R,2025PhRvD.111d3540G,2025JCAP...03..023A,2025arXiv250108915S,2025arXiv250114366S,2024arXiv241013627P,2024arXiv241212905C,2025EPJC...85..286O,2024arXiv240704385L,2024MNRAS.534.3869W,2024PhRvD.110l3512H,2024arXiv240408633C,2025PhRvD.112f3551G,2025arXiv250621010N,2026JHEAp..5000471O,2025JHEAp..4700398O,2025arXiv250416868A,2025ApJ...986L..31R,2024ApJ...976....1L,2025SCPMA..6910413L,2025JCAP...01..153J,2024PhRvD.110l3519J,2024PhRvD.110h3528W,2025PhRvD.111d1303W,2025JCAP...05..034W}. There are several types of instruments capable of measuring the CMB, including well-known space-based missions such as the currently most cited Planck satellite~\cite{planck2018-1,planck2018-2,planck2018-3,planckpr4,planckpr4lensing} and the now-retired Wilkinson Microwave Anisotropy Probe (WMAP)~\cite{wmap}, as well as ground-based observatories like the Atacama Cosmology Telescope (ACT)~\cite{actdr4,actdr6,actdr6-lensing-1,actdr6-lensing-2} and the South Pole Telescope (SPT)~\cite{spt-1,spt-2}. The CMB data provided by WMAP and Planck reach maximum multipoles of approximately 1200 and 2500 in temperature, and 800 and 2000 in polarization, respectively. In contrast, ground-based experiments such as ACT and SPT extend this coverage up to multipoles of 8500 and 3000, respectively, in both temperature and polarization. These ground-based experiments thus offer extended coverage in multipole space, enabling the exploration of smaller angular scales through high-resolution measurements of temperature and polarization anisotropies. However, they lack data at lower multipoles, which correspond to larger angular scales in the CMB power spectrum. The difference among these CMB datasets may lead to certain discrepancies in the resulting cosmological parameter constraints. For example, in the observational data used by the DESI collaboration to obtain results deviating from the $\Lambda$CDM, the CMB component includes the Planck \texttt{simall}, \texttt{Commander}, and \texttt{Plik} likelihoods, as well as the ACT-DR6 lensing likelihood~\cite{desidr2}. However, Ref.~\cite{William-CMB_cpl} claims that combining the same BAO and SNIa data with CMB data excluding Planck can reduce the observed deviation from $\Lambda$CDM. Moreover, the combination of DESI DR1 BAO data with non-Planck CMB data yields a more stringent constraint on the Hubble constant and reduces the significance of the Hubble tension~\cite{huang_H0_CMB}.

As mentioned above, when combining the same CMB and BAO data, different SNIa datasets lead to varying degrees of deviation from the $\Lambda$CDM cosmology, with Pantheon+ showing the least deviation and DESY5 the greatest. These variations highlight the importance of accurately calibrating the supernovae (SNe) distance scale at low to intermediate redshifts. As demonstrated in~\cite{2025MNRAS.538..875E}, a comparison of SNe common to both the DESY5 and Pantheon+ compilations reveals an offset of approximately 0.04 mag. between low and high redshifts, indicating the presence of systematic issues in the SN datasets. The authors argue that such unidentified systematics may bring the DESY5 sample into tension with the $\Lambda$CDM cosmology. However, Ref.~\cite{2025arXiv250106664V} provides an explanation for the origin of this offset and contends that the observed differences can be well understood. 

Of the 1829 SNe (DES-all) in the DESY5 dataset, 1635 (DES-SN) originate from the Dark Energy Survey (DES) survey program with homogeneous calibration and are almost all at redshifts greater than 0.1~\cite{DESY5-1,DESY5-2,DESY5-3}, while the remaining 194 low-redshift SNe (DES-lowz) are drawn from various historical observational programs with the best available calibration control~\cite{2017AJ....154..211K,2009ApJ...700..331H,2012ApJS..200...12H,2018MNRAS.475..193F}. Owing to its limited homogeneity and incomplete calibration, the DES-lowz sample represents the dominant source of systematics in the DESY5 dataset~\cite{DESY5-1,DESY5-2}. Therefore, the DESI collaboration excludes the low-redshift sample entirely and uses only the DES-SN sample, which is the most uniformly calibrated dataset. Naturally, this reduces the constraining power and, consequently, the statistical significance of the preference for dark energy evolution; however, the best-fit values of $w_0$ and $w_a$ remain significantly different from the $\Lambda$CDM cosmology~\cite{desidr2}. Ref.~\cite{2025SCPMA..6800413H} also investigates the impact of the DES-lowz sample on the deviation from the $\Lambda$CDM cosmology. Their analysis reveals substantial dispersion and a pronounced mismatch between the DES-lowz sample and the DES-SN sample within the DESY5 compilation. Accounting for these low-redshift systematics, with or without the inclusion of CMB data, significantly diminishes the statistical preference for dynamical dark energy to the $0.5 \sim 1.5 \sigma$ level.

For seven redshift bins in the range $0.1<z<4.2$, DESI BAO measurements provide high-precision constraints on the comoving angular diameter distance, the Hubble distance, and their combination. Several studies have identified the LRG1 and LRG2 BAO measurements as the primary drivers of the observed deviation from the $\Lambda$CDM among the seven redshift bins~\cite[e.g.][]{2024arXiv240704385L,2024MNRAS.534.3869W,2024PhRvD.110l3512H,2024arXiv240408633C,desidr1}. Moreover, there is a $\sim 3\sigma$ discrepancy between the comoving angular diameter distance of the LRG2 sample (with $z_{\rm eff} = 0.71$) in the DESI DR1 data and that of the LRG sample at $z = 0.7$ in the BAO measurements from the Sloan Digital Sky Survey (SDSS)~\cite{desidr1}. Additionally, the constraints on dynamical dark energy models derived from DESI BAO data show some differences compared to those obtained from the earlier SDSS BAO measurements~\cite[e.g.][]{2024arXiv240408633C,2024PhRvD.110l3533P,2024PDU....4601699G,2025JCAP...01..120D}. However, Ref.~\cite{desidr2} concludes that there is no significant discrepancy between the DESI DR2 measurements and those from SDSS. 

\begin{table}[ht]
\centering
% \footnotesize
\setlength{\tabcolsep}{2.5pt}
\renewcommand{\arraystretch}{1.2}  % 控制表格行高
% \hspace*{-0.8cm}
\begin{tabular}{|c|c|c|c|c|c|c|c|c|}
\hline
\textbf{Parameters} & $w_0$ & $w_a$ & $\Omega_bh^2$ & $\Omega_ch^2$ & $\theta_{MC}$ & $\tau$ & $\log(10^{10} A_\mathrm{s})$ & $n_s$  \\ \hline
\textbf{Prior} & $[-3,1]$&$[-3,2]$&$[0.005,1]$&$[0.01,0.99]$&$[0.005,0.1]$ &$[0.01,0.8]$&$[1.61,3.91]$&$[0.8,1.2]$ \\   \hline
\end{tabular}
\caption{\label{tab:para}The flat prior range of 8 free cosmological parameters constrained in this paper. As described in Section~\ref{sec:dam}, we impose a Gaussian prior on $\tau=0.0566\pm0.0058$ when performing analyses involving ACT, SPT and WMAP.}
\end{table}

In this paper, we examine the impact of different CMB datasets, various BAO measurements, and potential systematics in the DESY5 sample in order to investigate the origin of the deviations from the $\Lambda$CDM model. Our results show that the deviations are significantly reduced when considering datasets beyond Planck CMB, DESI DR2, and the DES-lowz sample. The remainder of this paper is organized as follows: In Section \ref{sec:dam}, we describe the datasets and methodology. Section \ref{sec:res} presents the results, while Section \ref{sec:dis} provides a discussion. Finally, Section \ref{sec:con} concludes the paper.

\section{Datasets and Methodology}
\label{sec:dam}
This study utilizes CMB measurements from WMAP, Planck, ACT, and SPT; BAO datasets from SDSS and DESI DR2; and the DESY5 SNIa sample. The details of these datasets are provided below.
\begin{itemize}
    \item WMAP: We use the nine-year CMB temperature and polarization data from WMAP~\cite{wmap}. The minimum multipole of the TE data is set to $l=24$ due to potential contamination by dust at lower multipoles. In addition, we include a Gaussian prior on $\tau=0.0566\pm0.0058$ (the mean value derived from the Planck \texttt{SRoll2} likelihood~\cite{sroll2} with a symmetric error) to account for the lack of low-$l$ data. 

    \item Planck: CMB temperature and polarization anisotropy power spectra (and their cross-spectra) from the Planck 2018 legacy data release (PR3) and NPIPE release in 2020 (PR4). For the low-$l$ data, we use \texttt{commander} likelihood for the TT spectrum and \texttt{SimAll} likelihood for the EE spectrum~\cite{planck2018-1,planck2018-2}, both in range $2\leq l \leq 29$; We use the NPIPE high-$l$ \texttt{CamSpec} likelihood for the TT spectrum in the multipole range $30\leq l \leq 2500$ and for the TE and EE spectra in the multipole range $30\leq l \leq 2000$~\cite{planckpr4highl-1,planckpr4highl-2}. We also add the Planck PR4 lensing likelihood in this work~\cite{planckpr4lensing}.

    \item ACT: We use measurements of CMB temperature, polarization, and lensing anisotropies from the Data Release 6 (DR6) maps produced by ACT, which include data collected from 2017 until the experiment concluded in 2022. Specifically, we use the \texttt{ACT-DR6} likelihood (covering the multipole range $600\leq l \leq 8500$)~\cite{actdr6} and the \texttt{ACT-DR6} lensing likelihood~\cite{actdr6-lensing-1,actdr6-lensing-2}, combined with a Gaussian prior on $\tau=0.0566\pm0.0058$. 

    \item SPT: We employ the CMB temperature and polarization (TT, TE, EE) anisotropy spectra provided by the SPT collaboration~\cite{spt-1,spt-2}. We also include a Gaussian prior on $\tau=0.0566\pm0.0058$. 

    \item SDSS: We use BAO measurements from SDSS, SDSS-II, the Baryon Oscillation Spectroscopic Survey (BOSS), and the extended BOSS (eBOSS), which comprising eight data points: one from the SDSS DR7 MGS sample~\cite{2015MNRAS.449..835R}, two from the BOSS DR12 galaxy sample~\cite{2016MNRAS.455.1553R}, and five from the eBOSS DR16 data~\cite{SDSS-BAO}. These measurements are summarized in Table III of Ref.~\cite{SDSS-BAO}.

    \item DESI: We use the latest DESI DR2 BAO measurements, include observations of galaxies, quasars, and Lyman-$\alpha$ tracers, as summarized in Table IV of Ref.~\cite{desidr2}.

\begin{table}[ht]
\centering
% \footnotesize
\setlength{\tabcolsep}{2.5pt}
\renewcommand{\arraystretch}{1.2}  % 控制表格行高
\begin{tabular}{|c|c|}
\hline
\textbf{Name} & \textbf{Likelihood} \\ \hline
\textbf{ACT-DR6} & \texttt{act\_dr6\_cmbonly\footnotemark[1]} \\ \hline
\textbf{ACT-DR6-Lensing} & \textbf{ACT-DR6} + \texttt{act\_dr6\_lenslike.ACTDR6LensLike\footnotemark[2]}\\ \hline
\multirow{3}{*}{\textbf{Planck}} & \texttt{planck\_2018\_lowl.TT\footnotemark[3]} \\ \cline{2-2}
\multirow{3}{*}{} & \texttt{planck\_2018\_lowl.EE\footnotemark[3]} \\ \cline{2-2}
\multirow{3}{*}{} & \texttt{planck\_NPIPE\_highl\_CamSpec.TTTEEE\footnotemark[3]} \\ \hline
\textbf{Planck-Lensing} & \textbf{Planck} + \texttt{planckpr4lensing.PlanckPR4Lensing\footnotemark[4]} \\ \hline
\textbf{SPT} & \texttt{spt3g\_2022.TTTEEE\footnotemark[5]} \\ \hline
\textbf{WMAP} & \texttt{wmaplike.WMAPLike\footnotemark[6]} \\ \hline
\multirow{7}{*}{\textbf{SDSS\,BAO}} & \texttt{bao.sdss\_dr7\_mgs\footnotemark[7]} \\ \cline{2-2}
\multirow{7}{*}{} & \texttt{bao.sdss\_dr12\_lrg\_bao\_dmdh\footnotemark[7]} \\ \cline{2-2}
\multirow{7}{*}{} & \texttt{bao.sdss\_dr16\_lrg\_bao\_dmdh\footnotemark[7]} \\ \cline{2-2}
\multirow{7}{*}{} & \texttt{bao.sdss\_dr16\_bao\_elg\footnotemark[7]} \\ \cline{2-2}
\multirow{7}{*}{} & \texttt{bao.sdss\_dr16\_qso\_bao\_dmdh\footnotemark[7]} \\ \cline{2-2}
\multirow{7}{*}{} & \texttt{bao.sdss\_dr16\_baoplus\_lyauto\footnotemark[7]} \\ \cline{2-2}
\multirow{7}{*}{} & \texttt{bao.sdss\_dr16\_baoplus\_lyxqso\footnotemark[7]} \\ \hline
\textbf{DESI\,DR2} & \texttt{bao.desi\_dr2\footnotemark[7]} \\ \hline
\end{tabular}
\caption{\label{tab:like}Likelihood functions of the CMB and BAO data implemented in \textbf{Cobaya} for different analyses.}
\end{table}

    \item DESY5: In this work, the DESY5 SNIa sample is divided into two parts. The low-redshift sample, DES-lowz, consists of 194 SNe, including 8 from the Carnegie Supernova Project~\cite{2017AJ....154..211K}, 68 from the Center for Astrophysics~\cite{2009ApJ...700..331H,2012ApJS..200...12H}, and 118 from the Pan-STARRS Supernova Survey~\cite{2018MNRAS.475..193F}. The DES-SN sample comprises 1,635 SNe collected during the full five years of the Dark Energy Survey Supernova Program~\cite{DESY5-1,DESY5-2,DESY5-3}.
\end{itemize}

\footnotetext[1]{\,\url{https://github.com/ACTCollaboration/DR6-ACT-lite/tree/main}}
\footnotetext[2]{\,\url{https://github.com/ACTCollaboration/act_dr6_lenslike}}
\footnotetext[3]{\,\url{https://cobaya.readthedocs.io/en/latest/likelihood_planck.html}}
\footnotetext[4]{\,\url{https://github.com/carronj/planck_PR4_lensing}}
\footnotetext[5]{\,\url{https://github.com/xgarrido/spt_likelihoods}}
\footnotetext[6]{\,\url{https://github.com/HTJense/pyWMAP}}
\footnotetext[7]{\,\url{https://cobaya.readthedocs.io/en/latest/likelihood_bao.html}}

We consider a cosmological model, $w_0w_a$CDM, where the dark energy equation of state is given by $w(a) = w_0 + w_a(1-a)$~\cite{cpl-1,cpl-2}. This model is characterized by eight parameters: the present-day value of the dark energy equation of state $w_0$ and $w_a$, describes the dynamical evolution of dark energy; the physical baryon density $\Omega_bh^2$ and physical cold dark matter density $\Omega_ch^2$; the angular size of the sound horizon $\theta_{MC}$; the optical depth to reionization $\tau$; the amplitude of primordial curvature perturbations $A_s$; and the scalar spectral index $n_s$. We use the public cosmological code \textbf{CAMB}~\cite{camb-1,camb-2} and the Markov Chain Monte Carlo (MCMC) sampler \textbf{Cobaya}~\cite{cobaya-1,cobaya-2,cobaya-3} to explore the posterior distributions of the eight-dimensional parameter space. The convergence of the MCMC chains is assessed using the Gelman–Rubin statistic~\cite{1992StaSc...7..457G}, requiring $|R-1| < 0.01$ as the tolerance criterion. The flat prior range of 8 free parameters is shown in Table~\ref{tab:para}. The likelihood functions of the CMB and BAO data, as implemented in \textbf{Cobaya} for the different analyses, are summarized in Table~\ref{tab:like}. Finally, we use \textbf{GetDist}~\cite{getdist} to perform the statistical analysis of the MCMC samples.

\begin{figure}
    \centering
    \includegraphics[width=1.0\linewidth]{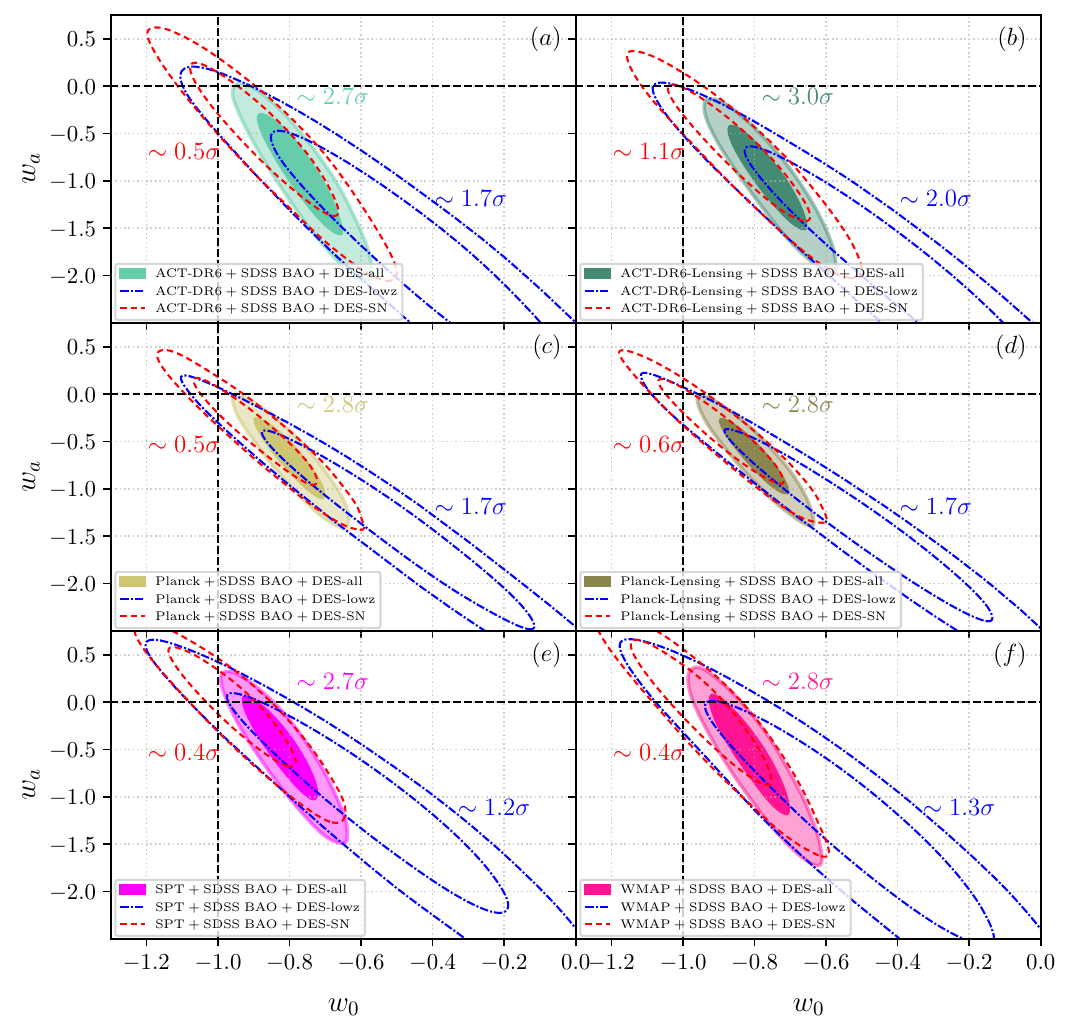}
    \caption{The marginalized $w_0$–$w_a$ distributions at 68\% and 95\% CL, obtained by combining different CMB data with SDSS BAO and DESY5 samples. In each panel, the red dashed lines and blue dash-dotted lines represent DES-SN and DES-lowz on the SNIa side, respectively. The black dashed lines indicate $w_0=-1$ and $w_a=0$ in the standard $\Lambda$CDM model.}
    \label{fig:eboss}
\end{figure}

\section{Results}
\label{sec:res}
\begin{figure}
    \centering
    \includegraphics[width=1.0\linewidth]{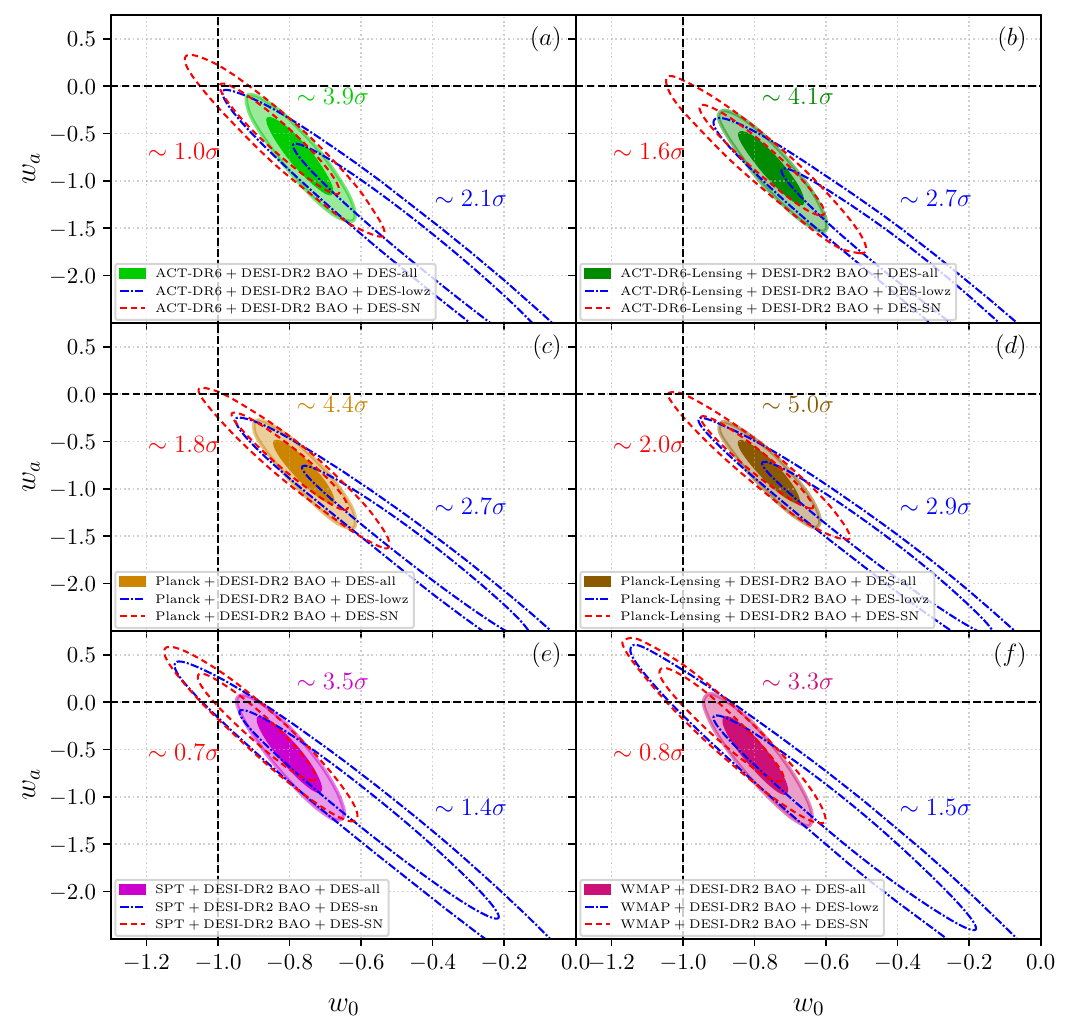}
    \caption{The same as in Figure~\ref{fig:eboss}, but with DESI-DR2 on the BAO side.}
    \label{fig:desi}
\end{figure}

In this study, we combine CMB, BAO, and SNIa data to constrain the $w_0w_a$CDM model, characterized by eight free parameters. We present the posterior distributions of all parameters, including derived parameters, from the different analyses in Tables~\ref{tab:act} to~\ref{tab:wmap} in Appendix~\ref{app}. The marginalized $w_0$–$w_a$ distributions are shown in Figures~\ref{fig:eboss} and~\ref{fig:desi}, highlighting the differences that arise from the choice of BAO data.

In Figure~\ref{fig:eboss}, we show the marginalized $w_0-w_a$ distribution constraints obtained from different CMB datasets in combination with SDSS BAO and DESY5 SNIa samples. We can see that even with different CMB datasets, using the DES-all sample on the SNIa side yields a significant deviation from $\Lambda$CDM at a level of at least $\sim 2.7\sigma$. In contrast, when only the DES-SN subset of the SNIa data is used, the point $w_0 = -1, w_a = 0$ lies well within the $1\sigma$ contour of the marginalized $w_0-w_a$ distributions. When only the DES-lowz subset is used, the deviation from $\Lambda$CDM is between $1\sigma$ and $2\sigma$; however, the uncertainties in the $w_0$–$w_a$ distributions increase significantly, rendering the deviation statistically insignificant. When using the CMB data from ACT-DR6 or Planck, we also tested whether including the lensing likelihood would affect the $w_0$–$w_a$ distributions. As shown in panel (a) and (b) in Figure~\ref{fig:eboss}, the \texttt{ACT-DR6} lensing likelihood has a small but non-negligible impact. The uncertainties of $w_0$ and $w_a$ change very little, but the mean value of $w_a$ decreases noticeably, making the results more inclined toward dynamical dark energy, even increasing the deviation when using DES-lowz from about $\sim 0.5\sigma$ to $\sim 1.1\sigma$. However, in panels (c) and (d) of Figure~\ref{fig:eboss}, including the Planck PR4 lensing likelihood with the Planck CMB data has almost no impact on the $w_0$–$w_a$ distributions. In the case of SDSS BAO, the systematics of the DESY5 sample play the most important role in the deviation from $\Lambda$CDM, consistent with the findings of Ref.~\cite{2025SCPMA..6800413H}, followed by the \texttt{ACT-DR6} lensing likelihood. The choice of CMB data has no significant impact, although using CMB datasets other than Planck in combination with SDSS BAO and DESY5 data results in relatively larger uncertainties in $w_0$ and $w_a$.

Then we replace the SDSS BAO data with DESI-DR2 BAO data, the marginalized $w_0$–$w_a$ distributions, shown in Figure~\ref{fig:desi}, exhibit differences from those obtained using SDSS BAO data. For the analyses using DES-all on the SNIa side, different CMB data lead to varying degrees of deviation from $\Lambda$CDM: WMAP yields a deviation of $\sim 3.3\sigma$, SPT $\sim 3.5\sigma$, ACT-DR6 $\sim 3.9\sigma$, and Planck $\sim 4.4\sigma$. Comparing panels (a) and (b), and (c) and (d) in Figure~\ref{fig:desi}, the \texttt{ACT-DR6} lensing likelihood and the Planck PR4 lensing likelihood both have a small but non-negligible impact, even pushing the Planck-Lensing+DESI DR2 BAO+DES-all combination to deviate from $\Lambda$CDM at the level of $\sim 5\sigma$. In each panel of Figure~\ref{fig:desi}, it is evident that using DES-SN on the SNIa side significantly reduces the deviation from $\Lambda$CDM, much like the effect of using SDSS on the BAO side. However, similar to the DES-all case, the choice of CMB dataset affects the degree of deviation from $\Lambda$CDM when DES-SN is used, with the deviations ranked as follows: Planck-Lensing$ > $Planck$ > $ACT-DR6-Lensing$ > $ACT-DR6$ > $WMAP$ > $SPT. Relative to the DES-all case, DES-lowz also mitigates the deviation from $\Lambda$CDM, but the uncertainties in the estimated values of $w_0$ and $w_a$ remain substantial. In the case of DESI-DR2 BAO, systematics in the DESY5 sample remain the dominant contributor to the deviation from the $\Lambda$CDM model, consistent with the findings of Ref.\cite{2025SCPMA..6800413H} and similar to results based on SDSS BAO data. The next most significant factor is the Planck CMB measurement, which also aligns with the conclusions of Ref.\cite{William-CMB_cpl}, suggesting that CMB experiments other than Planck tend to significantly reduce the deviation from $\Lambda$CDM.

Comparing the corresponding panels in Figures~\ref{fig:eboss} and \ref{fig:desi}, the DESI-DR2 BAO data lead to results that deviate more significantly from the $\Lambda$CDM. This is not solely due to the more precise DESI DR2 BAO measurements reducing the uncertainties on $w_0$ and $w_a$; rather, the mean values of these parameters are also shifted further away from $w_0=-1$ and $w_a=0$ compared to the SDSS BAO case. Consequently, the factors that most enhance the deviation from the $\Lambda$CDM are Planck on the CMB side, DESI-DR2 on the BAO side, and DES-lowz on the SNIa side.

\section{Discussion}
\label{sec:dis}
From the above results, we can see that using DES-SN on the SNIa side and SDSS on the BAO side, together with any CMB dataset, yields results consistent with $\Lambda$CDM within 1$\sigma$. When SDSS is replaced by DESI DR2 BAO data, combinations with SPT, WMAP, and ACT-DR6 on the CMB side remain consistent with $\Lambda$CDM within 1$\sigma$, while those with ACT-DR6-lensing, Planck, and Planck-lensing are consistent within 2$\sigma$. This indicates that we have identified combinations of CMB, BAO, and SNIa datasets that are consistent with $\Lambda$CDM. In the following, we analyze the possible causes of the deviations from $\Lambda$CDM among these three types of observations.

On the CMB side, the Planck data exhibit the most significant deviation from the $\Lambda$CDM model. This deviation is relatively modest when the SDSS BAO data are adopted but becomes substantially more pronounced with the DESI DR2 BAO data. As illustrated in panel (d) of Figure~\ref{fig:desi}, when employing the same BAO and SNIa datasets, the Planck-lensing results display the strongest departure from $\Lambda$CDM. One plausible explanation for this trend is that constraints derived from Planck CMB data have the smallest statistical uncertainties, which in turn magnifies any apparent deviation from the $\Lambda$CDM model. In addition, Planck delivers high-precision measurements at low multipoles (large angular scales), which are known to strengthen the preference for dynamical dark energy~\cite{William-CMB_cpl}. These low-$\ell$ modes ($\ell \lesssim 30$) are particularly sensitive to the Integrated Sachs–Wolfe (ISW) effect, and the most pronounced impact of dark-energy dynamics on the CMB power spectrum typically manifests as modifications to the ISW plateau amplitude~\cite{2024PhRvD.109j3519G}. Although WMAP also probes this regime, its lower sensitivity and higher noise lead to larger parameter uncertainties, thereby weakening the deviation from $\Lambda$CDM induced by the low-multipole information—yet it still shows a stronger preference for dynamical dark energy than the SPT datasets, which mainly constrain high-multipole modes that are largely insensitive to late-time dark-energy physics.

For the SNIa side, it is evident that, regardless of the dataset employed on the CMB or BAO side, using DES-all produces a strong deviation from $\Lambda$CDM, in agreement with the DESI collaboration’s previous constraints on dynamical dark energy \cite{desidr1,desidr2}. In contrast, adopting DES-SN on the SNIa side substantially reduces this deviation, most likely due to unknown systematics in the low-$z$ SN sample of the DESY5 compilation. As demonstrated in Ref.~\cite{2025SCPMA..6800413H}, the highly scattered DES-lowz sample shows a $\sim 0.043$ mag offset relative to DES-SN, consistent with Ref.~\cite{2025MNRAS.538..875E}, and this offset largely drives the apparent preference for dynamical dark energy. Compared to DES-all, using DES-SN reduces the deviation from $\Lambda$CDM by about $2$–$3\sigma$, not only because the absence of low-$z$ supernovae increases the uncertainties on $w_0$ and $w_a$, but also because the central values shift closer to the $\Lambda$CDM baseline ($w_0 = -1$, $w_a = 0$). Although the deviation from $\Lambda$CDM is also reduced when using the DES-lowz sample, this is primarily due to the significantly larger uncertainties in the $w_0$ and $w_a$ constraints, while the central values actually shift further away from the $\Lambda$CDM baseline compared to DES-all.

As discussed in the previous section, when using the same CMB and SNIa datasets, the DESI DR2 BAO results exhibit a larger deviation from the $\Lambda$CDM model than those based on SDSS BAO data. This discrepancy arises not only because the $w_0$ and $w_a$ constrained by the DESI DR2 BAO exhibit smaller uncertainties, but also because their best-fit parameter values deviate further from the $\Lambda$CDM baseline. This trend is clearly illustrated in the Tables~\ref{tab:act} to \ref{tab:wmap} provided in the Appendix~\ref{app}. According to Ref.~\cite{desidr2}, there is no significant discrepancy between the DESI DR2 BAO measurements and those from SDSS, and each individual redshift bin remains consistent within the $\Lambda$CDM framework. The largest internal difference is observed between the \texttt{LRG1} and \texttt{LRG3+ELG1} samples (see Fig.~7 in Ref.~\cite{desidr2}). Therefore, the larger deviation exhibited by the DESI DR2 BAO results may arise from the combined effects of their smaller statistical uncertainties and the mild tension among individual redshift bins. 

\section{Conclusion}
\label{sec:con}
The cosmological results released by the DESI collaboration deviate significantly from the $\Lambda$CDM, sparking considerable discussion and research interest. This has led to numerous studies aimed at identifying which component — CMB, BAO, or SNIa — may be responsible for the discrepancy. In this paper, we present a comprehensive examination of the impact of different CMB datasets, various BAO measurements, and potential systematics in the DESY5 sample. 

We find that using the DES-all SNIa sample consistently leads to a substantial deviation at more than $2.7\sigma$, regardless of the CMB and BAO dataset. In contrast, using only the DES-SN subset reduces the deviation to within the $1\sigma$ level, while the DES-lowz subset alone results in intermediate deviations but with large uncertainties. DES-all systematically favors dynamical dark energy, consistent with earlier DESI analyses. This preference is driven by unknown systematics in the DES-lowz sample, which shows a $\sim 0.043$ mag offset relative to DES-SN. Removing the low-$z$ sample (i.e., using DES-SN) not only enlarges the uncertainties of $(w_0, w_a)$ but also shifts their central values closer to $(w_0,w_a)=(-1,0)$, reducing the deviation from $\Lambda$CDM at the level of $2$–$3\sigma$. The ACT-DR6 lensing likelihood has a modest but noticeable impact, shifting the $w_a$ mean and slightly increasing the deviation from $\Lambda$CDM, whereas the Planck PR4 lensing likelihood has negligible influence. When replacing SDSS BAO data with DESI-DR2 BAO data, the deviation becomes not only because of tighter constraints, but also due to shifts in the best-fit values of $w_0$ and $w_a$. Among the different CMB datasets combined with DESI-DR2 BAO and DES-all, the deviation reaches up to $\sim 5\sigma$ with Planck and its lensing likelihood. This pronounced deviation is likely driven by Planck’s exceptionally small statistical uncertainties, as well as its high-precision measurements at low multipoles, which enhance sensitivity to late-time ISW effects and thereby amplify any underlying preference for dynamical dark energy. In contrast, WMAP’s larger noise weakens this effect, while SPT (dominated by high-$\ell$ modes) shows the smallest deviation because it is largely insensitive to dark‐energy dynamics. On the BAO side, DESI DR2 yields a larger deviation from $\Lambda$CDM than SDSS not only because its parameter uncertainties are smaller but also because its best-fit values lie further from the $\Lambda$CDM baseline. Although DESI DR2 is broadly consistent with SDSS and is internally consistent within the $\Lambda$CDM framework, mild tensions among its individual redshift bins, particularly between \texttt{LRG1} and \texttt{LRG3+ELG1}, together with its high statistical precision, collectively enhance the observed deviation. These findings emphasizes the importance of thoroughly understanding and mitigating potential systematics in SNIa samples, BAO measurements, and CMB analyses. The discrepancies between different data combination also point to the necessity of cross-checking results across independent experiments and data combinations to robustly assess the evidence for new physics beyond $\Lambda$CDM.

Future cosmological surveys and missions are expected to significantly improve the precision and control of systematics in each observational probe. The upcoming Vera C. Rubin Observatory (LSST)~\cite{lsst} will provide an unprecedented sample of well-calibrated SNIa spanning a wide redshift range, while the Zwicky Transient Facility (ZTF)~\cite{ztf} will deliver high-quality low-redshift SNIa samples. Together, these will help resolve the current inconsistencies among SNIa datasets. In addition, upcoming CMB observations (e.g. CMB-S4~\cite{cmbs4} and LiteBIRD~\cite{litebird}) and the continued progress in BAO measurements from DESI will offer deeper insights into validity of the dynamical dark energy.

\acknowledgments

We are grateful to Lin-Yu Li, Jia-Lei Niu, Hui-Qiang Liu, Shu-Yan Long and Yu-Xuan Li for their kind assistance and valuable discussions. This work was funded by the National Natural Science Foundation of China under Grant No. 12505070, No. 12375042, and No. 11975046, the Henan Provincial Natural Science Foundation No. 252300420902, the Startup Research Fund of Henan Academy of Sciences No. 241841222.

% Bibliography

%% [A] Recommended: using JHEP.bst file
\bibliographystyle{JHEP}
\bibliography{biblio.bib}

\newpage
\appendix
\section{Numerical constraints on cosmological parameters}
\label{app}
% Please always give a title also for appendices.
%table-actdr6
\begin{table}[ht]
\centering
\footnotesize
\setlength{\tabcolsep}{2pt}
\renewcommand{\arraystretch}{1.4}  % 控制表格行高
% \vspace*{-0.2cm}
\hspace*{-1.5cm}
\begin{tabular}{|c|cccccc|}
\hline
\multicolumn{1}{|c|}{\multirow{3}{*}{\textbf{Parameter}}} & \multicolumn{6}{c|}{\textbf{ACT-DR6+}}  \\ \cline{2-7} 
\multicolumn{1}{|c|}{}                  & \multicolumn{3}{c|}{\textbf{SDSS\,BAO+}}           & \multicolumn{3}{c|}{\textbf{DESI-DR2\,BAO+}}\\ \cline{2-7} 
\multicolumn{1}{|c|}{} & \multicolumn{1}{c|}{\textbf{DES-lowz}} & \multicolumn{1}{c|}{\textbf{DES-SN}} & \multicolumn{1}{c|}{\textbf{DES-all}} & \multicolumn{1}{c|}{\textbf{DES-lowz}} & \multicolumn{1}{c|}{\textbf{DES-SN}}  & \multicolumn{1}{c|}{\textbf{DES-all}} \\ \hline
$w_{0}$                      & \multicolumn{1}{c|}{$-0.46^{+0.27}_{-0.21}$}      & \multicolumn{1}{c|}{$-0.86^{+0.13}_{-0.15}$}   & \multicolumn{1}{c|}{$-0.765\pm 0.078$}            & \multicolumn{1}{c|}{$-0.43^{+0.24}_{-0.21}$}      & \multicolumn{1}{c|}{$-0.82^{+0.10}_{-0.12}$}      & $-0.769\pm 0.061$            \\ \hline
$w_{a}$                      & \multicolumn{1}{c|}{$-1.70^{+0.64}_{-0.88}$}      & \multicolumn{1}{c|}{$-0.59^{+0.61}_{-0.44}$}   & \multicolumn{1}{c|}{$-0.94^{+0.47}_{-0.37}$}      & \multicolumn{1}{c|}{$-1.70\pm 0.65$}              & \multicolumn{1}{c|}{$-0.58^{+0.42}_{-0.35}$}      & $-0.75^{+0.28}_{-0.25}$      \\ \hline
$\Omega_\mathrm{b}h^2$       & \multicolumn{1}{c|}{$0.02260\pm 0.00017$}         & \multicolumn{1}{c|}{$0.02260\pm 0.00017$}      & \multicolumn{1}{c|}{$0.02258\pm 0.00017$}         & \multicolumn{1}{c|}{$0.02260\pm 0.00017$}         & \multicolumn{1}{c|}{$0.02261\pm 0.00017$}         & $0.02261\pm 0.00017$         \\ \hline
$\Omega_\mathrm{c}h^2$       & \multicolumn{1}{c|}{$0.1221^{+0.0025}_{-0.0021}$} & \multicolumn{1}{c|}{$0.1214\pm 0.0025$}        & \multicolumn{1}{c|}{$0.1223^{+0.0025}_{-0.0022}$} & \multicolumn{1}{c|}{$0.1199^{+0.0016}_{-0.0013}$} & \multicolumn{1}{c|}{$0.1184^{+0.0017}_{-0.0014}$} & $0.1188^{+0.0015}_{-0.0013}$ \\ \hline
$\tau_\mathrm{reio}$         & \multicolumn{1}{c|}{$0.0563\pm 0.0042$}           & \multicolumn{1}{c|}{$0.0563\pm 0.0042$}        & \multicolumn{1}{c|}{$0.0567\pm 0.0058$}           & \multicolumn{1}{c|}{$0.0566\pm 0.0058$}           & \multicolumn{1}{c|}{$0.0566\pm 0.0058$}           & $0.0566\pm 0.0059$           \\ \hline
$\log(10^{10} A_\mathrm{s})$ & \multicolumn{1}{c|}{$3.058^{+0.042}_{-0.046}$}    & \multicolumn{1}{c|}{$3.072^{+0.043}_{-0.049}$} & \multicolumn{1}{c|}{$3.056\pm 0.044$}             & \multicolumn{1}{c|}{$3.080\pm 0.038$}             & \multicolumn{1}{c|}{$3.105\pm 0.040$}             & $3.098\pm 0.038$             \\ \hline
$n_\mathrm{s}$               & \multicolumn{1}{c|}{$0.9693\pm 0.0080$}           & \multicolumn{1}{c|}{$0.9707\pm 0.0082$}        & \multicolumn{1}{c|}{$0.9695\pm 0.0079$}           & \multicolumn{1}{c|}{$0.9741\pm 0.0073$}           & \multicolumn{1}{c|}{$0.9768\pm 0.0074$}           & $0.9761\pm 0.0073$           \\ \hline
$H_0$                        & \multicolumn{1}{c|}{$63.8^{+2.0}_{-2.4}$}         & \multicolumn{1}{c|}{$67.2\pm 1.1$}             & \multicolumn{1}{c|}{$66.67\pm 0.65$}              & \multicolumn{1}{c|}{$63.9^{+1.6}_{-2.1}$}         & \multicolumn{1}{c|}{$67.28\pm 0.99$}              & $66.89\pm 0.57$              \\ \hline
$r_{drag}$                   & \multicolumn{1}{c|}{$146.31^{+0.56}_{-0.64}$}     & \multicolumn{1}{c|}{$146.48\pm 0.65$}          & \multicolumn{1}{c|}{$146.28\pm 0.60$}             & \multicolumn{1}{c|}{$146.89^{+0.37}_{-0.43}$}     & \multicolumn{1}{c|}{$147.25^{+0.41}_{-0.46}$}     & $147.15\pm 0.41$             \\ \hline
$\Omega_m$                   & \multicolumn{1}{c|}{$0.358\pm 0.025$}             & \multicolumn{1}{c|}{$0.321\pm 0.012$}          & \multicolumn{1}{c|}{$0.3274\pm 0.0077$}           & \multicolumn{1}{c|}{$0.351\pm 0.022$}             & \multicolumn{1}{c|}{$0.313\pm 0.011$}             & $0.3176\pm 0.0060$           \\ \hline
             
Tension & \multicolumn{1}{c|}{$\sim 1.7 \sigma$} & \multicolumn{1}{c|}{$\sim 0.5 \sigma$} & \multicolumn{1}{c|}{$\sim 2.7 \sigma$ } & \multicolumn{1}{c|}{ $\sim 2.1 \sigma$} & \multicolumn{1}{c|}{ $\sim 1.0 \sigma$} & \multicolumn{1}{c|}{ $\sim 3.9 \sigma$} \\ \hline

\end{tabular}
\caption{\label{tab:act}Results at 68\% confidence level (CL) for the eight free cosmological parameters, three derived parameters and the tension with respect to the $\Lambda$CDM, obtained using ACT-DR6 in combination with two types of BAO data and DESY5 SNIa samples.}
\end{table}
%table-actdr6-lensing
\begin{table}[ht]
\centering
\footnotesize
\setlength{\tabcolsep}{2pt}
\renewcommand{\arraystretch}{1.4}  % 控制表格行高
% \vspace*{-0.5cm}
\hspace*{-1.5cm}
\begin{tabular}{|c|cccccc|}
\hline
\multicolumn{1}{|c|}{\multirow{3}{*}{\textbf{Parameter}}} & \multicolumn{6}{c|}{\textbf{ACT-DR6-Lensing+}}  \\ \cline{2-7} 
\multicolumn{1}{|c|}{}                  & \multicolumn{3}{c|}{\textbf{SDSS\,BAO+}}           & \multicolumn{3}{c|}{\textbf{DESI-DR2\,BAO+}}\\ \cline{2-7} 
\multicolumn{1}{|c|}{} & \multicolumn{1}{c|}{\textbf{DES-lowz}} & \multicolumn{1}{c|}{\textbf{DES-SN}} & \multicolumn{1}{c|}{\textbf{DES-all}} & \multicolumn{1}{c|}{\textbf{DES-lowz}} & \multicolumn{1}{c|}{\textbf{DES-SN}}  & \multicolumn{1}{c|}{\textbf{DES-all}} \\ \hline
$w_{0}$                      & \multicolumn{1}{c|}{$-0.44^{+0.27}_{-0.20}$} & \multicolumn{1}{c|}{$-0.83\pm 0.13$}         & \multicolumn{1}{c|}{$-0.759\pm 0.074$}       & \multicolumn{1}{c|}{$-0.38^{+0.23}_{-0.18}$}      & \multicolumn{1}{c|}{$-0.77^{+0.11}_{-0.12}$}      & $-0.750\pm 0.060$       \\ \hline
$w_{a}$                      & \multicolumn{1}{c|}{$-1.80^{+0.58}_{-0.84}$} & \multicolumn{1}{c|}{$-0.75^{+0.54}_{-0.42}$} & \multicolumn{1}{c|}{$-0.99^{+0.41}_{-0.33}$} & \multicolumn{1}{c|}{$-1.88^{+0.57}_{-0.69}$}      & \multicolumn{1}{c|}{$-0.80^{+0.42}_{-0.34}$}      & $-0.87^{+0.27}_{-0.24}$ \\ \hline
$\Omega_\mathrm{b}h^2$       & \multicolumn{1}{c|}{$0.02260\pm 0.00016$}    & \multicolumn{1}{c|}{$0.02259\pm 0.00016$}    & \multicolumn{1}{c|}{$0.02260\pm 0.00016$}    & \multicolumn{1}{c|}{$0.02258\pm 0.00016$}         & \multicolumn{1}{c|}{$0.02259\pm 0.00016$}         & $0.02259\pm 0.00016$    \\ \hline
$\Omega_\mathrm{c}h^2$       & \multicolumn{1}{c|}{$0.1225\pm 0.0018$}      & \multicolumn{1}{c|}{$0.1223\pm 0.0018$}      & \multicolumn{1}{c|}{$0.1225\pm 0.0018$}      & \multicolumn{1}{c|}{$0.1204^{+0.0013}_{-0.0011}$} & \multicolumn{1}{c|}{$0.1195^{+0.0014}_{-0.0012}$} & $0.1196\pm 0.0012$      \\ \hline
$\tau_\mathrm{reio}$         & \multicolumn{1}{c|}{$0.0571\pm 0.0057$}      & \multicolumn{1}{c|}{$0.0574\pm 0.0057$}      & \multicolumn{1}{c|}{$0.0571\pm 0.0057$}      & \multicolumn{1}{c|}{$0.0579\pm 0.0057$}           & \multicolumn{1}{c|}{$0.0585\pm 0.0057$}           & $0.0584\pm 0.0056$      \\ \hline
$\log(10^{10} A_\mathrm{s})$ & \multicolumn{1}{c|}{$3.038\pm 0.012$}        & \multicolumn{1}{c|}{$3.040\pm 0.012$}        & \multicolumn{1}{c|}{$3.039\pm 0.012$}        & \multicolumn{1}{c|}{$3.045\pm 0.011$}             & \multicolumn{1}{c|}{$3.049\pm 0.011$}             & $3.049\pm 0.011$        \\ \hline
$n_\mathrm{s}$               & \multicolumn{1}{c|}{$0.9678\pm 0.0074$}      & \multicolumn{1}{c|}{$0.9683\pm 0.0075$}      & \multicolumn{1}{c|}{$0.9677\pm 0.0074$}      & \multicolumn{1}{c|}{$0.9723\pm 0.0069$}           & \multicolumn{1}{c|}{$0.9740\pm 0.0070$}           & $0.9739\pm 0.0069$      \\ \hline
$H_0$                        & \multicolumn{1}{c|}{$63.8^{+1.9}_{-2.4}$}    & \multicolumn{1}{c|}{$67.2\pm 1.1$}           & \multicolumn{1}{c|}{$66.72\pm 0.64$}         & \multicolumn{1}{c|}{$63.6^{+1.6}_{-2.0}$}         & \multicolumn{1}{c|}{$67.1\pm 1.0$}                & $66.92\pm 0.56$         \\ \hline
$r_{drag}$                   & \multicolumn{1}{c|}{$146.21\pm 0.50$}        & \multicolumn{1}{c|}{$146.27\pm 0.51$}        & \multicolumn{1}{c|}{$146.19\pm 0.50$}        & \multicolumn{1}{c|}{$146.76\pm 0.36$}             & \multicolumn{1}{c|}{$147.00\pm 0.39$}             & $146.97\pm 0.36$        \\ \hline
$\Omega_m$                   & \multicolumn{1}{c|}{$0.359\pm 0.024$}        & \multicolumn{1}{c|}{$0.323\pm 0.011$}        & \multicolumn{1}{c|}{$0.3275\pm 0.0072$}      & \multicolumn{1}{c|}{$0.355^{+0.022}_{-0.020}$}    & \multicolumn{1}{c|}{$0.317\pm 0.011$}             & $0.3189\pm 0.0058$      \\ \hline

Tension & \multicolumn{1}{c|}{$\sim 2.0 \sigma$} & \multicolumn{1}{c|}{$\sim 1.1 \sigma$} & \multicolumn{1}{c|}{$\sim 3.0 \sigma$ } & \multicolumn{1}{c|}{ $\sim 2.7 \sigma$} & \multicolumn{1}{c|}{ $\sim 1.6 \sigma$} & \multicolumn{1}{c|}{ $\sim 4.1 \sigma$} \\ \hline

\end{tabular}
\caption{\label{tab:actl}Same with Table~\ref{tab:act}, but ACT-DR6 plus its lensing likelihood on CMB side.}
\end{table}
%table-planck
\begin{table}[ht]
\centering
\footnotesize
\setlength{\tabcolsep}{2pt}
\renewcommand{\arraystretch}{1.4}  % 控制表格行高
% \vspace*{-0.3cm}
\hspace*{-1.5cm}
\begin{tabular}{|c|cccccc|}
\hline
\multicolumn{1}{|c|}{\multirow{3}{*}{\textbf{Parameter}}} & \multicolumn{6}{c|}{\textbf{Planck+}}  \\ \cline{2-7} 
\multicolumn{1}{|c|}{}                  & \multicolumn{3}{c|}{\textbf{SDSS\,BAO+}}           & \multicolumn{3}{c|}{\textbf{DESI-DR2\,BAO+}}\\ \cline{2-7} 
\multicolumn{1}{|c|}{} & \multicolumn{1}{c|}{\textbf{DES-lowz}} & \multicolumn{1}{c|}{\textbf{DES-SN}} & \multicolumn{1}{c|}{\textbf{DES-all}} & \multicolumn{1}{c|}{\textbf{DES-lowz}} & \multicolumn{1}{c|}{\textbf{DES-SN}}  & \multicolumn{1}{c|}{\textbf{DES-all}} \\ \hline
$w_{0}$                      & \multicolumn{1}{c|}{$-0.49\pm 0.24$}         & \multicolumn{1}{c|}{$-0.88\pm 0.12$}         & \multicolumn{1}{c|}{$-0.795\pm 0.067$}       & \multicolumn{1}{c|}{$-0.44\pm 0.20$}      & \multicolumn{1}{c|}{$-0.79^{+0.10}_{-0.11}$} & $-0.757\pm 0.058$       \\ \hline
$w_{a}$                      & \multicolumn{1}{c|}{$-1.45^{+0.71}_{-0.64}$} & \multicolumn{1}{c|}{$-0.42^{+0.42}_{-0.34}$} & \multicolumn{1}{c|}{$-0.69^{+0.31}_{-0.26}$} & \multicolumn{1}{c|}{$-1.67\pm 0.58$}      & \multicolumn{1}{c|}{$-0.73^{+0.37}_{-0.30}$} & $-0.83^{+0.24}_{-0.21}$ \\ \hline
$\Omega_\mathrm{b}h^2$       & \multicolumn{1}{c|}{$0.02218\pm 0.00013$}    & \multicolumn{1}{c|}{$0.02218\pm 0.00013$}    & \multicolumn{1}{c|}{$0.02218\pm 0.00013$}    & \multicolumn{1}{c|}{$0.02220\pm 0.00013$} & \multicolumn{1}{c|}{$0.02223\pm 0.00013$}    & $0.02223\pm 0.00013$    \\ \hline
$\Omega_\mathrm{c}h^2$       & \multicolumn{1}{c|}{$0.1198\pm 0.0011$}      & \multicolumn{1}{c|}{$0.1197\pm 0.0011$}      & \multicolumn{1}{c|}{$0.1198\pm 0.0011$}      & \multicolumn{1}{c|}{$0.11946\pm 0.00092$} & \multicolumn{1}{c|}{$0.11894\pm 0.00092$}    & $0.11903\pm 0.00089$    \\ \hline
$\tau_\mathrm{reio}$         & \multicolumn{1}{c|}{$0.0509\pm 0.0075$}      & \multicolumn{1}{c|}{$0.0514\pm 0.0076$}      & \multicolumn{1}{c|}{$0.0511\pm 0.0075$}      & \multicolumn{1}{c|}{$0.0514\pm 0.0077$}   & \multicolumn{1}{c|}{$0.0523\pm 0.0078$}      & $0.0524\pm 0.0078$      \\ \hline
$\log(10^{10} A_\mathrm{s})$ & \multicolumn{1}{c|}{$3.034\pm 0.016$}        & \multicolumn{1}{c|}{$3.035\pm 0.016$}        & \multicolumn{1}{c|}{$3.034\pm 0.016$}        & \multicolumn{1}{c|}{$3.034\pm 0.016$}     & \multicolumn{1}{c|}{$3.035\pm 0.016$}        & $3.035\pm 0.016$        \\ \hline
$n_\mathrm{s}$               & \multicolumn{1}{c|}{$0.9631\pm 0.0038$}      & \multicolumn{1}{c|}{$0.9631\pm 0.0039$}      & \multicolumn{1}{c|}{$0.9629\pm 0.0039$}      & \multicolumn{1}{c|}{$0.9639\pm 0.0037$}   & \multicolumn{1}{c|}{$0.9650\pm 0.0036$}      & $0.9650\pm 0.0036$      \\ \hline
$H_0$                        & \multicolumn{1}{c|}{$63.5\pm 2.2$}           & \multicolumn{1}{c|}{$67.1\pm 1.0$}           & \multicolumn{1}{c|}{$66.42\pm 0.65$}         & \multicolumn{1}{c|}{$63.8^{+1.6}_{-2.0}$} & \multicolumn{1}{c|}{$67.0\pm 1.0$}           & $66.72\pm 0.57$         \\ \hline
$r_{drag}$                   & \multicolumn{1}{c|}{$147.37\pm 0.25$}        & \multicolumn{1}{c|}{$147.38\pm 0.25$}        & \multicolumn{1}{c|}{$147.36\pm 0.25$}        & \multicolumn{1}{c|}{$147.43\pm 0.23$}     & \multicolumn{1}{c|}{$147.54\pm 0.23$}        & $147.52\pm 0.22$        \\ \hline
$\Omega_m$                   & \multicolumn{1}{c|}{$0.355\pm 0.025$}        & \multicolumn{1}{c|}{$0.317\pm 0.010$}        & \multicolumn{1}{c|}{$0.3235\pm 0.0067$}      & \multicolumn{1}{c|}{$0.351\pm 0.020$}     & \multicolumn{1}{c|}{$0.3163\pm 0.0099$}      & $0.3188\pm 0.0057$      \\ \hline   
Tension & \multicolumn{1}{c|}{$\sim 1.7 \sigma$} & \multicolumn{1}{c|}{$\sim 0.5 \sigma$} & \multicolumn{1}{c|}{$\sim 2.8 \sigma$ } & \multicolumn{1}{c|}{ $\sim 2.7 \sigma$} & \multicolumn{1}{c|}{ $\sim 1.8 \sigma$} & \multicolumn{1}{c|}{ $\sim 4.4 \sigma$} \\ \hline

\end{tabular}
\caption{\label{tab:planck}Same with Table~\ref{tab:act}, but Planck on CMB side.}
\end{table}
%table-planck-lensing
\begin{table}[ht]
\centering
\footnotesize
\setlength{\tabcolsep}{2pt}
\renewcommand{\arraystretch}{1.4}  % 控制表格行高
% \vspace*{-0.2cm}
\hspace*{-1.5cm}
\begin{tabular}{|c|cccccc|}
\hline
\multicolumn{1}{|c|}{\multirow{3}{*}{\textbf{Parameter}}} & \multicolumn{6}{c|}{\textbf{Planck-Lensing+}}  \\ \cline{2-7} 
\multicolumn{1}{|c|}{}                  & \multicolumn{3}{c|}{\textbf{SDSS\,BAO+}}           & \multicolumn{3}{c|}{\textbf{DESI-DR2\,BAO+}}\\ \cline{2-7} 
\multicolumn{1}{|c|}{} & \multicolumn{1}{c|}{\textbf{DES-lowz}} & \multicolumn{1}{c|}{\textbf{DES-SN}} & \multicolumn{1}{c|}{\textbf{DES-all}} & \multicolumn{1}{c|}{\textbf{DES-lowz}} & \multicolumn{1}{c|}{\textbf{DES-SN}}  & \multicolumn{1}{c|}{\textbf{DES-all}} \\ \hline
$w_{0}$                      & \multicolumn{1}{c|}{$-0.50\pm 0.24$}      & \multicolumn{1}{c|}{$-0.88\pm 0.12$}         & \multicolumn{1}{c|}{$-0.797\pm 0.065$}       & \multicolumn{1}{c|}{$-0.45\pm 0.20$}         & \multicolumn{1}{c|}{$-0.79\pm 0.10$}         & $-0.757\pm 0.057$       \\ \hline
$w_{a}$                      & \multicolumn{1}{c|}{$-1.41\pm 0.65$}      & \multicolumn{1}{c|}{$-0.42^{+0.39}_{-0.34}$} & \multicolumn{1}{c|}{$-0.67^{+0.29}_{-0.25}$} & \multicolumn{1}{c|}{$-1.63^{+0.62}_{-0.55}$} & \multicolumn{1}{c|}{$-0.74^{+0.33}_{-0.30}$} & $-0.83^{+0.23}_{-0.21}$ \\ \hline
$\Omega_\mathrm{b}h^2$       & \multicolumn{1}{c|}{$0.02220\pm 0.00013$} & \multicolumn{1}{c|}{$0.02219\pm 0.00013$}    & \multicolumn{1}{c|}{$0.02219\pm 0.00013$}    & \multicolumn{1}{c|}{$0.02221\pm 0.00013$}    & \multicolumn{1}{c|}{$0.02224\pm 0.00013$}    & $0.02223\pm 0.00012$    \\ \hline
$\Omega_\mathrm{c}h^2$       & \multicolumn{1}{c|}{$0.11965\pm 0.00096$} & \multicolumn{1}{c|}{$0.11971\pm 0.00095$}    & \multicolumn{1}{c|}{$0.11972\pm 0.00096$}    & \multicolumn{1}{c|}{$0.11937\pm 0.00084$}    & \multicolumn{1}{c|}{$0.11894\pm 0.00082$}    & $0.11903\pm 0.00082$    \\ \hline
$\tau_\mathrm{reio}$         & \multicolumn{1}{c|}{$0.0506\pm 0.0071$}   & \multicolumn{1}{c|}{$0.0514\pm 0.0072$}      & \multicolumn{1}{c|}{$0.0509\pm 0.0072$}      & \multicolumn{1}{c|}{$0.0509\pm 0.0069$}      & \multicolumn{1}{c|}{$0.0530\pm 0.0070$}      & $0.0525\pm 0.0069$      \\ \hline
$\log(10^{10} A_\mathrm{s})$ & \multicolumn{1}{c|}{$3.033\pm 0.014$}     & \multicolumn{1}{c|}{$3.035\pm 0.014$}        & \multicolumn{1}{c|}{$3.034\pm 0.014$}        & \multicolumn{1}{c|}{$3.033\pm 0.014$}        & \multicolumn{1}{c|}{$3.037\pm 0.014$}        & $3.036\pm 0.014$        \\ \hline
$n_\mathrm{s}$               & \multicolumn{1}{c|}{$0.9639\pm 0.0038$}   & \multicolumn{1}{c|}{$0.9635\pm 0.0039$}      & \multicolumn{1}{c|}{$0.9635\pm 0.0038$}      & \multicolumn{1}{c|}{$0.9645\pm 0.0036$}      & \multicolumn{1}{c|}{$0.9655\pm 0.0037$}      & $0.9653\pm 0.0036$      \\ \hline
$H_0$                        & \multicolumn{1}{c|}{$63.5^{+2.0}_{-2.4}$} & \multicolumn{1}{c|}{$67.1\pm 1.0$}           & \multicolumn{1}{c|}{$66.40\pm 0.64$}         & \multicolumn{1}{c|}{$63.9^{+1.7}_{-1.9}$}    & \multicolumn{1}{c|}{$67.00\pm 0.97$}         & $66.72\pm 0.56$         \\ \hline
$r_{drag}$                   & \multicolumn{1}{c|}{$147.39\pm 0.23$}     & \multicolumn{1}{c|}{$147.38\pm 0.23$}        & \multicolumn{1}{c|}{$147.38\pm 0.23$}        & \multicolumn{1}{c|}{$147.45\pm 0.21$}        & \multicolumn{1}{c|}{$147.52\pm 0.21$}        & $147.51\pm 0.21$        \\ \hline
$\Omega_m$                   & \multicolumn{1}{c|}{$0.355\pm 0.025$}     & \multicolumn{1}{c|}{$0.317\pm 0.010$}        & \multicolumn{1}{c|}{$0.3234\pm 0.0066$}      & \multicolumn{1}{c|}{$0.350\pm 0.020$}        & \multicolumn{1}{c|}{$0.3161\pm 0.0096$}      & $0.3188\pm 0.0056$      \\ \hline

Tension & \multicolumn{1}{c|}{$\sim 1.7 \sigma$} & \multicolumn{1}{c|}{$\sim 0.6 \sigma$} & \multicolumn{1}{c|}{$\sim 2.8 \sigma$ } & \multicolumn{1}{c|}{ $\sim 2.9 \sigma$} & \multicolumn{1}{c|}{ $\sim 2.0 \sigma$} & \multicolumn{1}{c|}{ $\sim 5.0 \sigma$} \\ \hline

\end{tabular}
\caption{\label{tab:planckl}Same with Table~\ref{tab:act}, but Planck plus its lensing likelihood on CMB side.}
\end{table}
%table-spt
\begin{table}[ht]
\centering
\footnotesize
\setlength{\tabcolsep}{2pt}
\renewcommand{\arraystretch}{1.4}  % 控制表格行高
% \vspace*{-0.3cm}
\hspace*{-1.5cm}
\begin{tabular}{|c|cccccc|}
\hline
\multicolumn{1}{|c|}{\multirow{3}{*}{\textbf{Parameter}}} & \multicolumn{6}{c|}{\textbf{SPT+}}  \\ \cline{2-7} 
\multicolumn{1}{|c|}{}                  & \multicolumn{3}{c|}{\textbf{SDSS\,BAO+}}           & \multicolumn{3}{c|}{\textbf{DESI-DR2\,BAO+}}\\ \cline{2-7} 
\multicolumn{1}{|c|}{} & \multicolumn{1}{c|}{\textbf{DES-lowz}} & \multicolumn{1}{c|}{\textbf{DES-SN}} & \multicolumn{1}{c|}{\textbf{DES-all}} & \multicolumn{1}{c|}{\textbf{DES-lowz}} & \multicolumn{1}{c|}{\textbf{DES-SN}}  & \multicolumn{1}{c|}{\textbf{DES-all}} \\ \hline
$w_{0}$                      & \multicolumn{1}{c|}{$-0.56\pm 0.25$}         & \multicolumn{1}{c|}{$-0.94^{+0.11}_{-0.13}$}  & \multicolumn{1}{c|}{$-0.818^{+0.066}_{-0.076}$} & \multicolumn{1}{c|}{$-0.57^{+0.22}_{-0.25}$}      & \multicolumn{1}{c|}{$-0.89^{+0.10}_{-0.12}$}      & $-0.797\pm 0.060$            \\ \hline
$w_{a}$                      & \multicolumn{1}{c|}{$-1.13^{+0.83}_{-0.69}$} & \multicolumn{1}{c|}{$-0.099^{+0.50}_{-0.34}$} & \multicolumn{1}{c|}{$-0.51^{+0.42}_{-0.31}$}    & \multicolumn{1}{c|}{$-1.21^{+0.79}_{-0.64}$}      & \multicolumn{1}{c|}{$-0.28^{+0.41}_{-0.34}$}      & $-0.56^{+0.28}_{-0.25}$      \\ \hline
$\Omega_\mathrm{b}h^2$       & \multicolumn{1}{c|}{$0.02226\pm 0.00031$}    & \multicolumn{1}{c|}{$0.02227\pm 0.00031$}     & \multicolumn{1}{c|}{$0.02226\pm 0.00031$}       & \multicolumn{1}{c|}{$0.02223\pm 0.00031$}         & \multicolumn{1}{c|}{$0.02225\pm 0.00031$}         & $0.02225\pm 0.00031$         \\ \hline
$\Omega_\mathrm{c}h^2$       & \multicolumn{1}{c|}{$0.1177\pm 0.0027$}      & \multicolumn{1}{c|}{$0.1170\pm 0.0027$}       & \multicolumn{1}{c|}{$0.1180\pm 0.0026$}         & \multicolumn{1}{c|}{$0.1176^{+0.0020}_{-0.0016}$} & \multicolumn{1}{c|}{$0.1160^{+0.0019}_{-0.0017}$} & $0.1167^{+0.0017}_{-0.0015}$ \\ \hline
$\tau_\mathrm{reio}$         & \multicolumn{1}{c|}{$0.0554\pm 0.0058$}      & \multicolumn{1}{c|}{$0.0557\pm 0.0058$}       & \multicolumn{1}{c|}{$0.0555\pm 0.0057$}         & \multicolumn{1}{c|}{$0.0555\pm 0.0057$}           & \multicolumn{1}{c|}{$0.0558\pm 0.0058$}           & $0.0557\pm 0.0058$           \\ \hline
$\log(10^{10} A_\mathrm{s})$ & \multicolumn{1}{c|}{$3.042\pm 0.018$}        & \multicolumn{1}{c|}{$3.042\pm 0.018$}         & \multicolumn{1}{c|}{$3.043\pm 0.018$}           & \multicolumn{1}{c|}{$3.041\pm 0.018$}             & \multicolumn{1}{c|}{$3.039\pm 0.018$}             & $3.040\pm 0.018$             \\ \hline
$n_\mathrm{s}$               & \multicolumn{1}{c|}{$0.968\pm 0.015$}        & \multicolumn{1}{c|}{$0.969\pm 0.016$}         & \multicolumn{1}{c|}{$0.967\pm 0.015$}           & \multicolumn{1}{c|}{$0.969\pm 0.015$}             & \multicolumn{1}{c|}{$0.971\pm 0.015$}             & $0.970\pm 0.015$             \\ \hline
$H_0$                        & \multicolumn{1}{c|}{$63.5^{+2.1}_{-2.5}$}    & \multicolumn{1}{c|}{$67.0\pm 1.1$}            & \multicolumn{1}{c|}{$66.21\pm 0.68$}            & \multicolumn{1}{c|}{$64.5\pm 2.0$}                & \multicolumn{1}{c|}{$67.2\pm 1.0$}                & $66.54\pm 0.58$              \\ \hline
$r_{drag}$                   & \multicolumn{1}{c|}{$147.83\pm 0.79$}        & \multicolumn{1}{c|}{$148.03\pm 0.80$}         & \multicolumn{1}{c|}{$147.77\pm 0.76$}           & \multicolumn{1}{c|}{$147.91^{+0.53}_{-0.60}$}     & \multicolumn{1}{c|}{$148.31\pm 0.59$}             & $148.11\pm 0.55$             \\ \hline
$\Omega_m$                   & \multicolumn{1}{c|}{$0.350\pm 0.026$}        & \multicolumn{1}{c|}{$0.312\pm 0.011$}         & \multicolumn{1}{c|}{$0.3214\pm 0.0077$}         & \multicolumn{1}{c|}{$0.339^{+0.022}_{-0.025}$}    & \multicolumn{1}{c|}{$0.307\pm 0.011$}             & $0.3154\pm 0.0060$           \\ \hline
Tension & \multicolumn{1}{c|}{$\sim 1.2 \sigma$} & \multicolumn{1}{c|}{$\sim 0.4 \sigma$} & \multicolumn{1}{c|}{$\sim 2.7 \sigma$ } & \multicolumn{1}{c|}{ $\sim 1.4 \sigma$} & \multicolumn{1}{c|}{ $\sim 0.7 \sigma$} & \multicolumn{1}{c|}{ $\sim 3.5 \sigma$} \\ \hline

\end{tabular}
\caption{\label{tab:spt}Same with Table~\ref{tab:act}, but SPT on CMB side.}
\end{table}
%table-wmap
\begin{table}[ht]
\centering
\footnotesize
\setlength{\tabcolsep}{2pt}
\renewcommand{\arraystretch}{1.4}  % 控制表格行高
% \vspace*{-0.3cm}
\hspace*{-1.5cm}
\begin{tabular}{|c|cccccc|}
\hline
\multicolumn{1}{|c|}{\multirow{3}{*}{\textbf{Parameter}}} & \multicolumn{6}{c|}{\textbf{WMAP+}}  \\ \cline{2-7} 
\multicolumn{1}{|c|}{}                  & \multicolumn{3}{c|}{\textbf{SDSS\,BAO+}}           & \multicolumn{3}{c|}{\textbf{DESI-DR2\,BAO+}}\\ \cline{2-7} 
\multicolumn{1}{|c|}{} & \multicolumn{1}{c|}{\textbf{DES-lowz}} & \multicolumn{1}{c|}{\textbf{DES-SN}} & \multicolumn{1}{c|}{\textbf{DES-all}} & \multicolumn{1}{c|}{\textbf{DES-lowz}} & \multicolumn{1}{c|}{\textbf{DES-SN}}  & \multicolumn{1}{c|}{\textbf{DES-all}} \\ \hline
$w_{0}$                      & \multicolumn{1}{c|}{$-0.53\pm 0.25$}              & \multicolumn{1}{c|}{$-0.93^{+0.12}_{-0.14}$}      & \multicolumn{1}{c|}{$-0.807^{+0.069}_{-0.082}$}   & \multicolumn{1}{c|}{$-0.55\pm 0.24$}              & \multicolumn{1}{c|}{$-0.89\pm 0.11$}              & $-0.795^{+0.057}_{-0.064}$ \\ \hline
$w_{a}$                      & \multicolumn{1}{c|}{$-1.27\pm 0.78$}              & \multicolumn{1}{c|}{$-0.14^{+0.59}_{-0.41}$}      & \multicolumn{1}{c|}{$-0.58^{+0.48}_{-0.35}$}      & \multicolumn{1}{c|}{$-1.26\pm 0.74$}              & \multicolumn{1}{c|}{$-0.25^{+0.43}_{-0.36}$}      & $-0.57^{+0.29}_{-0.25}$    \\ \hline
$\Omega_\mathrm{b}h^2$       & \multicolumn{1}{c|}{$0.02234\pm 0.00044$}         & \multicolumn{1}{c|}{$0.02239\pm 0.00045$}         & \multicolumn{1}{c|}{$0.02231\pm 0.00044$}         & \multicolumn{1}{c|}{$0.02236\pm 0.00043$}         & \multicolumn{1}{c|}{$0.02245\pm 0.00043$}         & $0.02240\pm 0.00043$       \\ \hline
$\Omega_\mathrm{c}h^2$       & \multicolumn{1}{c|}{$0.1177^{+0.0035}_{-0.0030}$} & \multicolumn{1}{c|}{$0.1164^{+0.0038}_{-0.0033}$} & \multicolumn{1}{c|}{$0.1178^{+0.0033}_{-0.0030}$} & \multicolumn{1}{c|}{$0.1173^{+0.0027}_{-0.0024}$} & \multicolumn{1}{c|}{$0.1154^{+0.0028}_{-0.0024}$} & $0.1165\pm 0.0024$         \\ \hline
$\tau_\mathrm{reio}$         & \multicolumn{1}{c|}{$0.0562\pm 0.0058$}           & \multicolumn{1}{c|}{$0.0565\pm 0.0057$}           & \multicolumn{1}{c|}{$0.0563\pm 0.0059$}           & \multicolumn{1}{c|}{$0.0565\pm 0.0058$}           & \multicolumn{1}{c|}{$0.0564\pm 0.0057$}           & $0.0564\pm 0.0058$         \\ \hline
$\log(10^{10} A_\mathrm{s})$ & \multicolumn{1}{c|}{$3.040\pm 0.018$}             & \multicolumn{1}{c|}{$3.037\pm 0.018$}             & \multicolumn{1}{c|}{$3.040\pm 0.018$}             & \multicolumn{1}{c|}{$3.040\pm 0.017$}             & \multicolumn{1}{c|}{$3.034\pm 0.017$}             & $3.037\pm 0.017$           \\ \hline
$n_\mathrm{s}$               & \multicolumn{1}{c|}{$0.964\pm 0.011$}             & \multicolumn{1}{c|}{$0.966\pm 0.011$}             & \multicolumn{1}{c|}{$0.963\pm 0.011$}             & \multicolumn{1}{c|}{$0.965\pm 0.010$}             & \multicolumn{1}{c|}{$0.9681\pm 0.0098$}           & $0.966\pm 0.010$           \\ \hline
$H_0$                        & \multicolumn{1}{c|}{$63.4^{+2.1}_{-2.4}$}         & \multicolumn{1}{c|}{$67.0\pm 1.1$}                & \multicolumn{1}{c|}{$66.18\pm 0.74$}              & \multicolumn{1}{c|}{$64.4^{+1.9}_{-2.2}$}         & \multicolumn{1}{c|}{$67.3\pm 1.1$}                & $66.60\pm 0.69$            \\ \hline
$r_{drag}$                   & \multicolumn{1}{c|}{$147.8\pm 1.0$}               & \multicolumn{1}{c|}{$148.0\pm 1.1$}               & \multicolumn{1}{c|}{$147.77\pm 0.99$}             & \multicolumn{1}{c|}{$147.84\pm 0.93$}             & \multicolumn{1}{c|}{$148.26\pm 0.93$}             & $148.02\pm 0.92$           \\ \hline
$\Omega_m$                   & \multicolumn{1}{c|}{$0.351\pm 0.025$}             & \multicolumn{1}{c|}{$0.311\pm 0.013$}             & \multicolumn{1}{c|}{$0.3214\pm 0.0083$}           & \multicolumn{1}{c|}{$0.339\pm 0.024$}             & \multicolumn{1}{c|}{$0.306\pm 0.011$}             & $0.3146\pm 0.0063$         \\ \hline

Tension & \multicolumn{1}{c|}{$\sim 1.3 \sigma$} & \multicolumn{1}{c|}{$\sim 0.4 \sigma$} & \multicolumn{1}{c|}{$\sim 2.8 \sigma$ } & \multicolumn{1}{c|}{ $\sim 1.5 \sigma$} & \multicolumn{1}{c|}{ $\sim 0.8 \sigma$} & \multicolumn{1}{c|}{ $\sim 3.3 \sigma$} \\ \hline

\end{tabular}
\caption{\label{tab:wmap}Same with Table~\ref{tab:act}, but WMAP on CMB side.}
\end{table}

% \paragraph{Note added.} This is also a good position for notes added
% after the paper has been written.

%% or
%% [B] Manual formatting (see below)
%% (i) We suggest to always provide author, title and journal data or doi:
%% in short all the informations that clearly identify a document.
%% (ii) please avoid comments such as "For a review'', "For some examples",
%% "and references therein" or move them in the text. In general, please leave only references in the bibliography and move all
%% accessory text in footnotes.
%% (iii) Also, please have only one work for each \bibitem.

% \begin{thebibliography}{99}

% \bibitem{a}
% Author,
% \emph{Title},
% \emph{J. Abbrev.} {\bf vol} (year) pg.

% \bibitem{b}
% Author,
% \emph{Title},
% arxiv:1234.5678.

% \bibitem{c}
% Author,
% \emph{Title},
% Publisher (year).

% \end{thebibliography}
\end{document}